# Boron-10 lined RPCs for sub-millimeter resolution thermal neutron detectors: Conceptual design and performance considerations

**L. M. S. Margato,** [a,1] **A. Morozov**[a]

[a] *LIP-Coimbra, Departamento de Física, Universidade de Coimbra*
   *Rua Larga, 3004-516 Coimbra, Portugal*

ABSTRACT: A novel concept for position-sensitive thermal neutron detectors based on thin-gap Resistive Plate Chambers (RPC) lined with a neutron converter layer is presented and its feasibility is discussed. Several detector architectures implementing a stack of RPCs in multilayer and inclined geometries are introduced. The results of a Monte Carlo simulation study are presented demonstrating the effect of the main detector design parameters on the expected detection efficiency and spatial resolution.

---

[1] E-mail: margato@coimbra.lip.pt

1. **Introduction**

Experiments in Neutron Scattering Science (NSS), planned for the future high-flux neutron sources such as, e.g., the European Spallation Source (ESS), establish new requirements for thermal neutron position sensitive detectors (PSD) which are still unreachable with the state-of-the-art detection technologies based on $^3$He alternatives [1-4]. This motivates an active development of a new generation of thermal neutron PSDs with improved characteristics, which can also benefit practical applications in, e.g., homeland security and geology.

Invented in the eighties of the last century [5], Resistive Plate Chambers (RPC) have been widely used in large-area detectors (>100 m$^2$) for experiments in high energy and astroparticle physics [6-10] due to good resolution in both time and position [11]. Further, a possibility to detect slow neutrons with RPCs has already been suggested before [12]. However, despite the attractive properties of RPCs, such as modular detector designs, good scalability and low cost per unit area, development of RPCs-based detectors for thermal neutron imaging applications in NSS and homeland security has not been pursued. RPCs have neither been applied in high resolution and high detection efficiency PSDs for thermal neutrons. The only reported application of this technology was developed for finding antipersonnel land mines based on detection of back-scattered neutrons from hydrogen-rich explosive materials [13,14].

By itself, RPCs are essentially insensitive to thermal neutrons. A neutron converter lined onto the surfaces of the RPC plates facing the gas-gap is required to have a neutron capture reaction followed by the emission of charged particles. Only a very limited set of isotopes can effectively accomplish this role, e.g. $^3$He, $^6$Li, $^{10}$B and $^{157}$Gd. Not considering the scarce and expensive $^3$He [15,16], $^{10}$B in the form of a solid material, such as B$_4$C, appears to be the most promising candidate [17].

One of the main challenges associated with $^{10}$B solid converters is a very low (~5%) maximum detection efficiency that can be achieved with a single layer for normal neutron incidence compared to, for example, nearly 100% efficiency of $^3$He proportional counters. Two main approaches have been developed to overcome this limitation. One is to use multiple converter layers, and the other is to position the converters at a small angle with respect to the direction of incoming neutrons [18]. Both approaches have been implemented in detectors based on wire counters [19-22] and gas electron multipliers (GEM) [23,24].

Due to their layered structure and high modularity, RPCs are well suited for both approaches. Furthermore, compared to the other multilayer detectors based on multi-wires or GEMs, a multilayer RPC detector has an advantage: for the same number of converter layers, the amount of material exposed to the neutron beam is smaller, resulting in lower neutron scattering.

While providing promising results in terms of detection efficiency, the currently available detectors with $^{10}$B converters offer intrinsic spatial resolution of about 3 mm FWHM (see, e.g., [24]), which is significantly worse than the 0.5 mm resolution achieved in the past with $^3$He-based detectors [25,26]. We expect that $^{10}$B-lined thin-gap RPCs ($^{10}$B-RPC) can reach sub-millimeter spatial resolutions with a potential to surpass 100 μm achieved for RPCs with minimum ionizing particles (MIPs) [11]. This estimation is based on the *anti-parallax* effect of the electron avalanche development in RPCs described below and on the possibility to build RPCs with very thin gas-gaps (down to 0.1 mm), allowing to confine the primary ionization to the close vicinity of the neutron capture position by considerably shortening the ranges of the fission fragments in the gas.

Another attractive feature of RPCs is their fast timing [11], enabling to measure the neutrons time-of-flight (TOF) with nanosecond resolution. This makes detectors based on $^{10}$B-RPC a promising tool for experiments profiting from energy-selective or fast dynamics neutron imaging.

In this paper a conceptual design of a position-sensitive neutron detector (PSND) based on $^{10}$B-RPCs is described and several detector architectures aiming at high neutron detection efficiency (>50%) and sub-millimeter spatial resolution are introduced. The main characteristics expected for this type of detectors are also discussed based on numerical simulations.

## 2. PSNDs based on $^{10}$B-lined RPCs

### 2.1 Basic concept

The structure of a $^{10}$B-RPC is similar to a standard RPC [5], except that one of its electrode plates is coated on the side facing the gas-gap with a thin layer of a solid converter containing $^{10}$B. A sketch of such an RPC is shown in figure 1. The sensitivity to thermal neutrons originates from the $^{10}$B (n, α) $^7$Li capture reaction:

$$^{10}\text{B} + n \rightarrow {}^7\text{Li}\,(0.84\text{MeV}) + {}^4\text{He}\,(1.47\text{MeV}) + \gamma\,(0.48\text{MeV}) \quad (94\%)$$
$$^{10}\text{B} + n \rightarrow {}^7\text{Li}\,(1.02\text{MeV}) + {}^4\text{He}\,(1.78\text{MeV}) \quad (6\%)$$

The fission fragments ($^4$He and $^7$Li) are isotropically emitted in opposite directions. If the converter layer is thin enough, one of the fragments can exit into the gas-gap and generate primary ionization. Under high-strength electric field the primary ionization electrons initiate a Townsend avalanche process, which, in turn, induces signals in the pick-up electrodes.

The converter layer can be deposited on the cathode as shown in figure 1, on the anode plate, or on both of them. However, deposition only on the cathode seems to be the best option. The fission fragments which have dissipated the most of their energy in the converter generate primary ionization only in the vicinity of the corresponding electrode. In this case, the cathode side has an advantage: the primary ionization electrons have a longer path for multiplication and, therefore, can form larger avalanches, which results in higher probability for the neutron to be detected.

The second consideration favouring deposition of the converter on the cathodes results from what can be called the *anti-parallax* effect. The fission fragments exiting the converter and entering deeply into the gas-gap with large angles from the normal direction to the plates generate long ionization tracks, resulting in a non-negligible offset between the positions of the neutron capture and those of the primary ionization clusters. The avalanches triggered by electrons in the vicinity of the cathode (in the beginning of the track, and, hence, close to the neutron capture position) have longer multiplication paths and become larger, thereby having a greater contribution to the induced signals. The centroid of the induced charge is thus located closer to the neutron capture position compared to the scenario where all electron clusters along the track lead to avalanches of a similar size, resulting in an improved spatial resolution.

In the case when the converter is lined onto the anode, the largest avalanches are triggered by the clusters of electrons far from the neutron capture position leading to a much higher uncertainty in the reconstructed neutron positions.

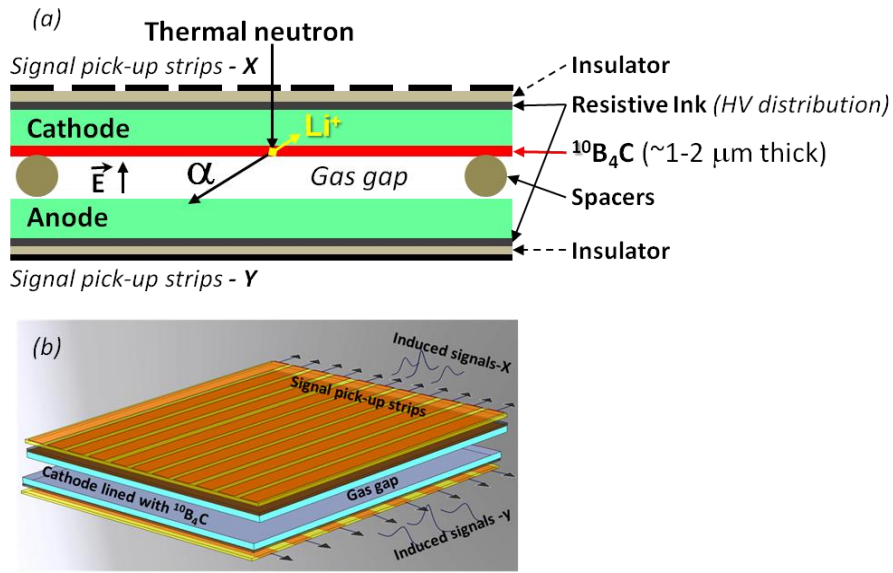

**Figure 1.** Cross-section (a) and a 3D view (b) of the conceptual design of $^{10}$B-RPCs. The cathode is lined on the gas-gap side with a thin layer of $^{10}$B$_4$C.

The cathode and anode plates have to be assembled parallel to each other to form a gas-gap of a constant width in order to ensure uniformity of the electric field over the active area thus avoiding fluctuations in the detection efficiency. This is typically achieved by means of insulating spacers as shown in figure 1. Using this method, RPCs operating with gas-gaps widths down to 0.1 mm have already been reported [27].

In order to preserve the spark protection mechanism [28], at least one of the RPC plates have to be made of a material with high volume resistivity ($10^8$–$10^{13}$ $\Omega\cdot$cm). Typically, glass, phenolic laminates or ceramics are used for this purpose (see [27,29,30] and the references therein).

The surfaces of the resistive plates on the side opposite to the gas-gap have to be coated with a resistive layer, made of, e.g., a graphite compound or an acrylic ink. This layer is used to uniformly distribute the high voltage (HV) over the entire surface of the plate. The resistive layers can be produced using, for example, spray-coating or screen-printing techniques [31]. Their surface resistivity should be in the range of $10^6$ - $10^8$ $\Omega/\square$ in order to be transparent to the induced signals, which is needed to preserve the event localization capability [32].

To establish a strong electric field (several tens of kV/cm) inside the gas-gap, the HV can be applied symmetrically to the anode (+HV) and the cathode (-HV) or, if it is more convenient from the practical considerations, to only one of the electrodes, keeping the other one at the ground potential.

The composition of the working gas can vary substantially depending on the chosen operation mode (streamer or avalanche) of the RPCs [28,33]. We expect the streamer mode to be less suited for neutron detectors due to the higher sensitivity of the RPCs to gamma rays and the lower maximum counting rate capability.

Calculations performed with SRIM code [34,35] suggest that the specific energy loss of $^4$He and $^7$Li particles in Tetrafluoroethane ($C_3H_2F_4$), a gas commonly used in RPCs, is about three orders of magnitude higher than that for MIPs [36]. Therefore, the fission fragments should generate a number of electron-ion pairs about 2 - 3 orders of magnitude higher.

Consequently, it is expected that the RPC efficiency plateau for the $^4$He and $^7$Li particles should be significantly shifted towards the lower voltages, allowing to operate the detector in a regime with low dark noise and low sensitivity to gamma rays (and MIPs in general). This feature is especially important since there is no coincidence trigger to discriminate non-neutron events.

For practical reasons the working gas is typically kept at the atmospheric pressure and circulated in an open loop mode. However, detector operation at lower or higher gas pressures can be considered in order to adjust the range of the charged particles inside the gas-gap.

The 2D position of the neutron capture can be determined, for instance, using the signals [37] induced in two mutually orthogonal arrays of strips, one facing the anode and the other facing the cathode as shown in figure 1. Both X and Y coordinates of an event can be estimated by applying, for example the Center of Gravity (CoG) algorithm [38].

As shown in figure 1, the strips are typically decoupled from the HV on the resistive layers by a thin insulator (e.g. polyamide film). To avoid significant neutron scattering in hydrogen-rich polymers, the electrical insulation can also be achieved using, e.g., a very thin film of $Al_2O_3$ or $SiO_2$ made by physical or chemical vapor deposition.

The signal pickup electrodes can also be designed as an independent structure, mechanically decoupled from the RPC. This results in a high level of modularity, simplifying both the assembly and maintenance of the detector.

Other position reconstruction methods, requiring a smaller number of electronic channels, such as, e.g., resistive and capacitive charge-division [39] or delay line techniques [40,41] can also be implemented for RPCs.

## 2.2 Hybrid RPCs

The $^{10}$B-RPC configuration discussed in the previous section has two technological challenges making their realization difficult. The first one is the problem of achieving long-term adhesion stability of the converter layer onto the resistive substrate (e.g. float glass). The second one is tailoring the surface resistivity of the converter to a large enough value ($>10^6$ $\Omega/\square$) in order to avoid shielding of the induced signals in the pick-up electrodes.

Both issues can be avoided by using the so-called *hybrid RPC* design which allows to combine one resistive and one metallic plate [28,42]. Using this approach, $^{10}$B converters can be deposited on the metallic cathodes (see figure 2). This is much more straightforward to accomplish since deposition of $^{10}B_4C$ on aluminium substrates is already a well-established technique [17,43].

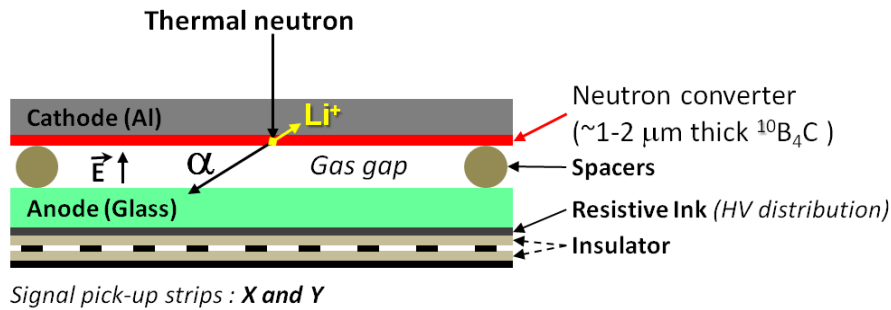

**Figure 2.** Conceptual design of a $^{10}$B-lined hybrid RPC. The metallic cathode, made of, e.g., aluminium, is coated on the side facing the gas-gap with the neutron converter layer.

The 2D position readout is, however, more challenging for the hybrid configuration: since the cathode is metallic, pickup of signals for both X and Y coordinates have to be made on the resistive anode side. This demands a more intricate design of the signal pickup electrodes. One possibility is to use two adjacent arrays of parallel and mutually orthogonal strips, manufactured, for example, on a thin double-layered PCB. The strips facing the anode should have a suitable width and pitch to allow sharing of the induced signals with the second array of strips, situated behind. An RPC configuration, which is even more convenient for practical applications can be derived from the hybrid RPC design by combining the RPC plates in such a way that a metallic cathode is shared by two adjacent resistive anodes, defining two gas-gaps as shown in figure 3. In this case the cathode is lined with a neutron converter on both sides. We refer to this configuration below as the *$^{10}$B double-gap RPC*.

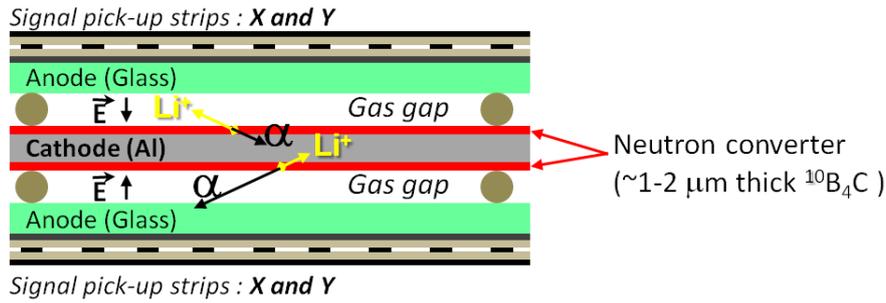

**Figure 3.** Double-gap RPC configuration: a single metallic cathode, shared by two anodes, is lined on both sides with a $^{10}$B neutron converter.

## 2.3 Architectures for high detection efficiency

The maximum detection efficiency, which can be reached with a single layer of $^{10}B_4C$ converter for normally-incident thermal neutrons strongly depends on the neutron wavelength and is typically limited to the approximate range $\approx$ 2 - 10%. This limitation originates from the fact that in order to detect a neutron capture event, one of the fission fragments ($^7$Li or $^4$He), generated inside the converter, has to reach the gas-gap and deposit there a sufficient amount of energy. Since the ranges of these particles in $^{10}B_4C$ are short ($\approx$1.8 μm for $^7$Li at 0.84 MeV and $\approx$3.5 μm for $^4$He at 1.47 MeV), the converter layer has to be quite thin (1 - 2 μm). On the other hand, in order to achieve capture probability greater than 80%, neutrons with λ = 1.8 Å have to have a path longer than 40 μm inside the converter. A compromise between a high neutron capture probability and a short escape distance for the fission fragments can be achieved either by stacking several $^{10}$B-RPCs, or by placing them at a small angle with respect to the neutron beam direction.

### 2.3.1 Multilayer configuration

A significant improvement in the detection efficiency, compared to the basic design described in section 2.2, can be achieved by stacking several $^{10}$B-RPCs. The conceptual design of such detector is shown in figure 4.

Stacking $^{10}$B double-gap RPCs has several advantages compared to stacking single-gap ones. It reduces the material budget per neutron converter and significantly simplifies electrical insulation: HV can be applied to the cathodes, keeping the anodes, which face the signal pickup

– 6 –

electrodes, at the ground potential. As shown in figure 4, the pickup electrodes can be shared by two neighbouring RPCs thus reducing the number of readout structures.

Signals from the cathodes can be used to identify the triggered RPC making possible to determine the neutron capture coordinate along the beam direction and, thus, to measure the neutron time-of-flight (TOF). Another advantage of a multilayer detector is that the maximum counting rate capability increases with the number of RPCs in the stack.

The multilayer architecture allows to build detectors with a highly modular design, using two types of independent building blocks: the blocks of double-gap RPCs and the blocks of signal pickup electrodes. This is a very convenient feature for mass-production and maintenance of the detectors.

The main drawback shared by all multilayer configurations is a higher gamma sensitivity and neutron scattering due to increased material budget compared to the single-layer detectors.

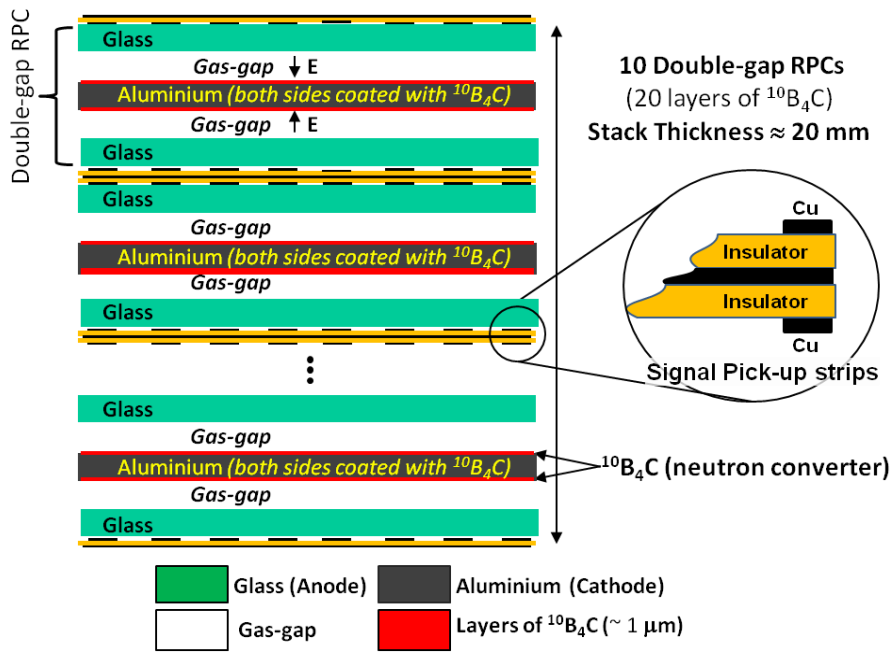

**Figure 4.** Conceptual design of a PSND with $^{10}$B double-gap RPCs in a multilayer architecture implementing a stack of double-gap RPCs.

2.3.2 **Multigap RPCs**

A Multigap RPC (MRPC) [44] is another type of multilayer architecture that can be used to improve the detection efficiency (figure 5). It consists of a stack of several resistive plates defining several gas-gaps. All the inner plates of the stack are at the floating potential and the external surfaces of the first and the last plates are coated with a resistive layer to distribute HV, as discussed in section 2.1. The plates, made of a resistive material ($10^8 - 10^{12}$ $\Omega$·cm) such as, e.g., float glass, should be lined with a $^{10}$B converter on the cathode sides (see discussion in section 2.1).

– 7 –

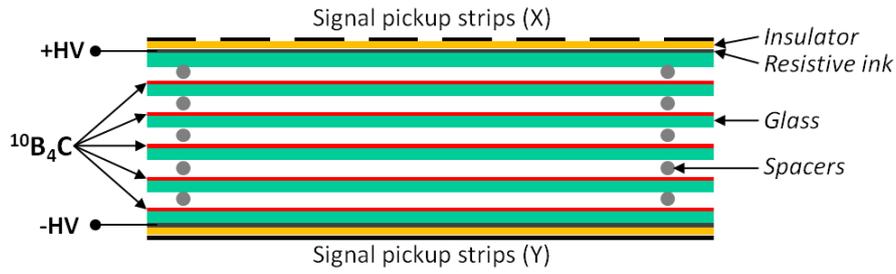

**Figure 5.** Conceptual design of a $^{10}$B-lined MRPC.

The main advantage of MRPCs is that the signal pickup electrodes are shared by all the gas-gaps. This significantly reduces the overall number of electrodes needed for the position readout, which decreases the material budget of the detector. Another advantage of this approach is the possibility to create a very thin detector: for example, an MRPC with ten $^{10}$B$_4$C converters can be only 5 to 10 mm thick.

The main challenges of MRPC-based PSNDs are ensuring long-term stable adhesion of $^{10}$B converters to the resistive plates and tuning their surface resistivity to a high enough value ($> 10^7$ Ω/□). The latter is needed to avoid a broad spread of the induced charge, which hinders the determination of the neutron capture position.

### 2.3.3 RPCs at inclined geometries

Another approach to improve the detection efficiency is to use the detector design where $^{10}$B-RPCs are placed at a small angle $\theta$ (typically 5º - 10º) with respect to the direction of the incoming neutrons [18,22]. As shown in figure 6, the effective thickness of $^{10}$B$_4$C layers crossed by the neutrons is increased by a factor of $1/sin(\theta)$. This allows to significantly enhance the neutron capture probability while maintaining high the chance for fission fragments to exit the converter into the gas-gap.

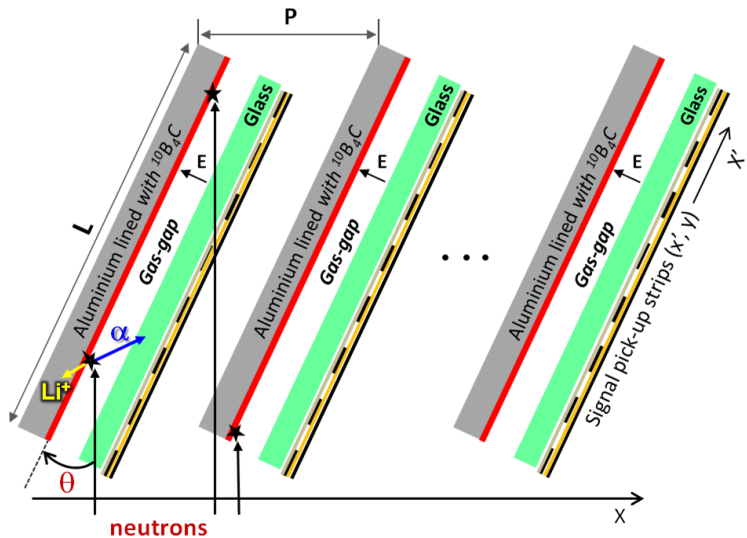

**Figure 6.** Conceptual design of a PSND implementing several single-gap $^{10}$B-RPCs at an inclined geometry.

Figure 6 shows a PSND design with single-gap $^{10}$B-RPCs arranged at an inclined geometry. The position readout can be organized in the same way as described in section 2.1. Note that for the X-direction the spatial resolution is improved by a factor of $1/sin(\theta)$ relative to the non-inclined configuration: for example, assuming a value of 0.5 mm for the intrinsic resolution of the RPCs, the expected spatial resolution in the X-direction for the angle of 5º is better than 50 μm. Also, due to the distribution of the neutron flux over a $1/sin(\theta)$ larger area, the maximum counting rate should be higher by the same factor in comparison to the non-inclined single-gap RPC.

The pitch $P$ at which $^{10}$B-RPCs have to be installed depends on the angle $\theta$ and on the length $L$ of the RPCs plates: $P \leq L \cdot sin(\theta)$. The fact that the RPC length can be extended without a negative effect on its performance can be used to decrease the required number of RPCs.

For detectors using the configuration in which the fission fragments exit the converter layer into the gas-gap in the backward direction (opposite to the incoming neutrons, see figure 6), it is advantageous to have thick converter layers (> 3μm). In this case essentially all neutrons are captured in the first converter on their path, thus minimizing the contribution of the fluctuations of the converter thickness to the non-uniformity of the detection efficiency and reducing the neutron scattering.

The main challenge associated with the inclined geometry is the non-uniformity in the detection efficiency [45] due to the edge effects at the transitions between the successive $^{10}$B-RPCs.

## 3. MC simulations: design and performance considerations

In this section we discuss the effect of several parameters of the detector design on the expected performance characteristics, mainly on detection efficiency and spatial resolution. The discussion is based on the results of a MC simulation study and is limited to the most important parameters, such as the number of converter layers, their thickness, the width of the gas-gap, and, for the inclined geometries, the angle between the converter layers and the direction of incidence of neutrons.

It is assumed that the neutron converters are made of $^{10}$B$_4$C, which currently appears to be the most practical material [17,43]. For $^{10}$B$_4$C the neutron scattering cross section is three orders of magnitude lower than that of capture and thus scattering in the converter layers can be ignored. For the RPC plates made of float glass and aluminium (both 0.5 mm thick) and 1 μm thick $^{10}$B$_4$C converter layers, simulations show that the scattering probability is one order of magnitude lower than the probability of neutron capture in the converter. Since the contribution of scattering to attenuation of the neutron beam is weak and depends strongly on the geometry and material composition of a particular detector we show here only the results of the simulations performed for the ideal case by ignoring the neutron scattering. Thus only absorption of thermal neutrons by the detector materials and the $^{10}$B (n, α) $^7$Li capture reaction in the neutron converter layers were taken into account.

A neutron is considered detected if one of the reaction products ($^7$Li or $^4$He) exits the neutron converter and deposits in the gas-gap an energy of at least 10 keV. All simulations were performed using the ANTS2 v4.08 toolkit [46].

## 3.1 Simulation parameters

The $^{10}B_4C$ neutron converters (composition: B:81.7 + C:17 + H:0.7 + O:0.4 + N:0.2) were assumed to have a high (95%) enrichment in $^{10}B$ and the density of 2.24 g/cm$^3$. This isotopic composition is taken from [17] where it was determined by time-of-flight elastic recoil detection analysis (ToF-ERDA).

Following the discussion in section 2.1, the material for the RPC electrodes was chosen to be float glass (composition: $SiO_2$:72.98 + $Na_2O$:14 + CaO:7 + MgO:4 + $Al_2O_3$:2 + $K_2O$:0.02) with a density of 2.53 g/cm$^3$. It was assumed that the RPCs gas-gaps are filled with Tetrafluoroethane ($C_2H_2F_4$) at standard conditions (20 °C and 1 atm).

For all elements except Boron, the natural isotope abundances were used. The energy-resolved neutron absorption cross-sections were obtained from the Nuclear Data Services database of the International Atomic Energy Agency [47]. The main source of data was the ENDF/B-VII.1 library, however, if cross-section data for a given isotope were missing, the JEFF-3.2 and JENDL-4.0 libraries were used instead. The stopping power for $^7Li$ and $^4He$ particles in the neutron converters and in the gas was calculated using the SRIM package [35]. To provide sufficient statistics, all simulations were performed with at least $2\times10^5$ incident neutrons.

## 3.2 Neutron converter thickness

As discussed in section 2.3, the detection efficiency is largely limited by the short ranges of the neutron capture reaction products inside the neutron converter. Overcoming this limitation by using the multilayer approach (section 2.3.1) requires a careful optimization of the neutron converter thickness. The optimal thickness depends mainly on the wavelength (or on the distribution of the wavelengths) of the incident neutrons and on the number of the converter layers to be implemented in the detector.

For the detection efficiency calculations we have chosen a detector design based on the simplest and the most general single-gap RPC geometry shown in figure 1. The gas-gap width was 0.35 mm, and the anodes and cathodes were made of 0.35 mm thick float glass. The surface of the cathodes facing the gas-gap was lined with a layer of $^{10}B_4C$. A multilayer configuration was generated by stacking a certain number (N = 1 .. 20) of the single-gap $^{10}B$-RPCs.

The detection efficiency was obtained as a function of the converter layer thickness for different number of layers and considering several neutron wavelengths (see figure 7). The converter thickness was the same for all RPCs in the stack. The neutron beam was normal to the RPC plates and irradiated the RPCs in the forward direction ($^7Li$ and $^4He$ particles exit to the gas-gap in the direction of the neutron beam).

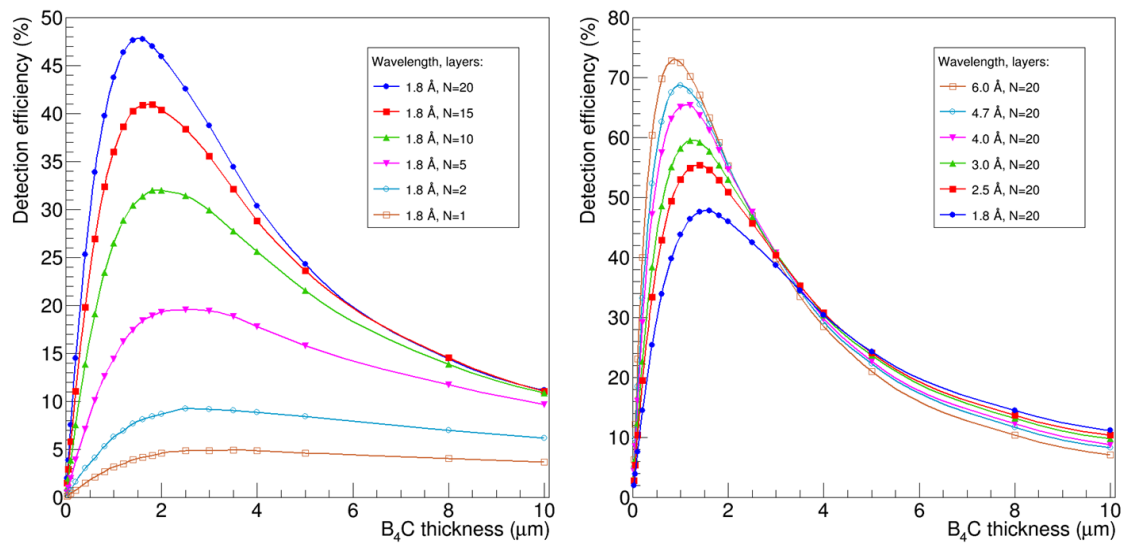

**Figure 7.** Detection efficiency as a function of the $^{10}B_4C$ thickness. **Left:** Neutron wavelength is 1.8 Å, the number of layers is ranging from 1 to 20. **Right:** 20 converter layers, the neutron wavelength is ranging from 1.8 Å to 6 Å.

The simulation results show that the larger is the number of converter layers, the smaller is the optimal thickness of the $^{10}B_4C$ converters (see figure 7, left). For example, increasing the number of layers from 5 to 20, the optimum converter thickness reduces from ≈ 2.5 μm to ≈ 1.5 μm. Moreover, with increase of the number of converter layers the detector generally becomes more sensitive to the fluctuations in the converter thickness as can be concluded based on the sharpening of the peak in figure 7 (left).

A similar study performed for a fixed number of converters shows that the optimal thickness reduces with the neutron wavelength (see figure 7, right). The results also demonstrate that the range of optimal thickness becomes somewhat narrower with the increase of the wavelength.

It is also shown in figure 7 (right) that the maximum detection efficiency increases with the neutron wavelength, which is explained by the proportionality of the neutron capture cross section to the inverse square root of the neutron energy. Overall, the obtained result indicates that a detection efficiency higher than 50% can be achieved.

The next step in optimization of the multilayer configuration which can be performed for a particular detector is to find individual thicknesses of the converter layers targeting not only the highest overall detection efficiency, but also equalizing the counting rates in all RPCs of the stack as much as possible. In comparison to the detector having all converter layers with the same thicknesses, the second condition allows to improve the counting rate capability of the detector by avoiding much earlier saturation of the first RPCs in the stack.

### 3.3 Gas-gap width

A series of simulations using configurations different only by the width of the gas-gap were performed to analyze the influence of this parameter on the performance of $^{10}B$-RPCs. For the widths of 0.35, 1 and 2 mm we have obtained the distributions of the projected ranges of the capture reaction fragments in the gas along the direction parallel to the RPC planes, and the distributions of the energy deposited in the gas by these particles. The simulations were performed for 1.8 Å neutrons and assuming a converter layer thickness of 2 µm.

The results shown in figure 8 demonstrate that the gas-gap width has a significant effect on the lateral range of the fission fragments in the gas. The wider is the gas-gap, the larger is their range, and the broader is the spread of the primary ionization. Therefore, $^{10}$B-RPCs with thinner gas-gap widths have a potential to reach smaller uncertainties in the reconstructed neutron capture positions.

As shown in figure 9, wider gas-gaps also lead, in average, to a larger energy deposition and a broader distribution of the energy. A larger energy deposition may allow better gamma discrimination, but a broader distribution results in a larger range of the signal amplitudes, which, in turn, requires front-end electronics with a higher dynamic range. For wider gas-gaps, the probability of converting gamma rays into primary ionization is higher, which may result in higher gamma sensitivity. Wider gas-gaps also require higher bias voltages, making electrical insulation more demanding. We consider a gas-gap width of 0.35 mm to be a good compromise, taking into account that RPC detectors with these gas-gaps have already been successfully operated [11].

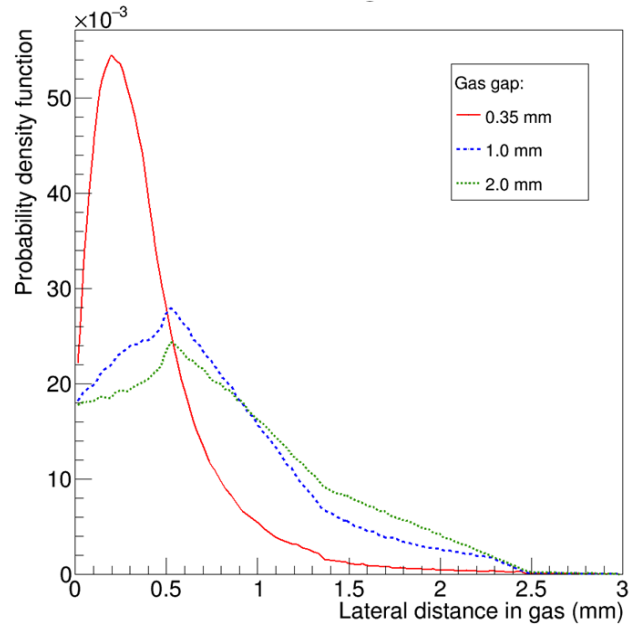

**Figure 8.** Distributions of the $^4$He and $^7$Li particles ranges in the gas-gap, projected in the lateral direction (parallel to the RPC plates) for several gas-gap widths (0.35, 1.0 and 2 mm). The gas-gap is filled with $C_2H_2F_4$ (20ºC and 1 atm), the $^{10}B_4C$ thickness is 2 µm and the neutron wavelength is 1.8 Å.

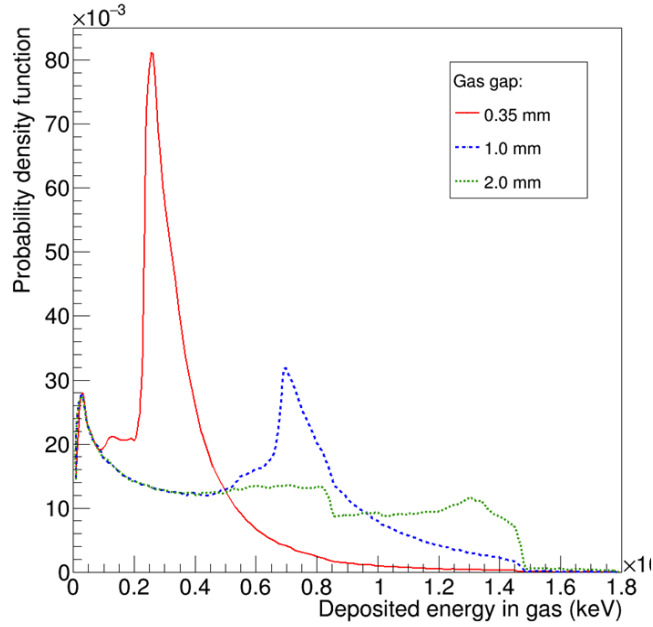

**Figure 9.** Deposited energy by the $^4$He and $^7$Li particles in the gas-gap filled with $C_2H_2F_4$ (20ºC and 1 atm) for several gas-gap widths (0.35, 1.0 and 2 mm). $^{10}B_4C$ thickness is 2 μm and the neutron wavelength is 1.8 Å.

### 3.4 Angle of incidence

In this section we discuss the results of simulations for the inclined detector architecture, introduced in section 2.3.3. The RPCs are placed at a small angle to the neutron beam, which irradiates the RPCs in the backward direction in respect to the neutron converter (see figure 6).

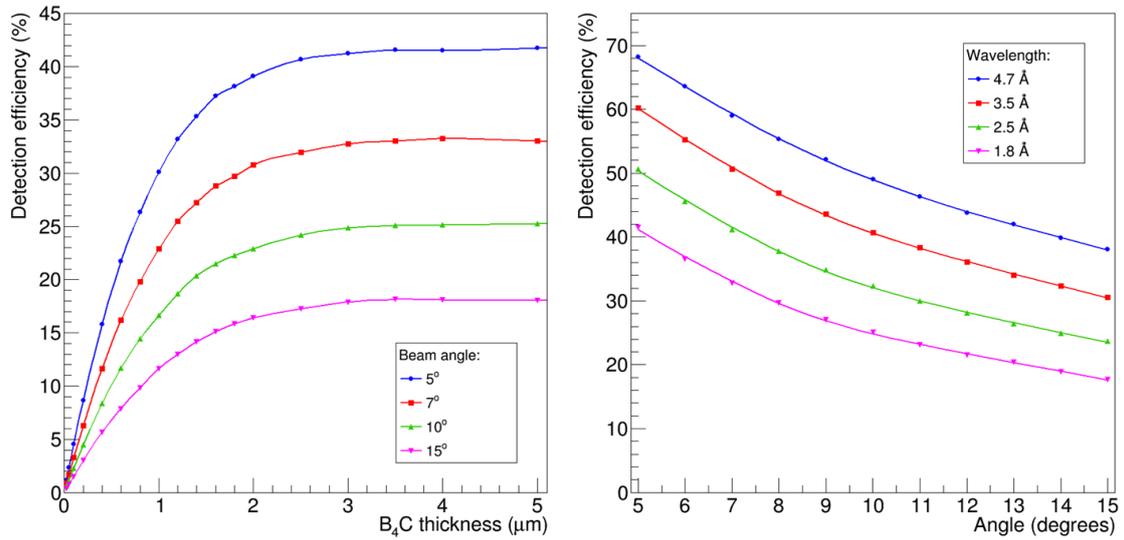

**Figure 10. Left:** Simulated detection efficiency as a function of the $^{10}B_4C$ thickness for several angles of incidence of the neutron beam; **Right:** detection efficiency as a function of the angle of incidence of the beam for several neutron wavelengths and a $^{10}B_4C$ thickness of 3 μm.

In the first series of simulations the converter thickness was ranging from 0.05 to 5 μm, the neutron wavelength was fixed to 1.8 Å and several angles of neutron incidence were used (5º, 7º, 10º and 15º). The results presented in figure 10 (left) show that for a converter thickness above ≈3 μm the detection efficiency is essentially constant. Therefore, fluctuations in the converter thickness from the deposition process should have essentially no impact on the detection uniformity if the thickness is larger than this value.

As expected, the simulations confirm that smaller angles lead to a higher efficiency due to the longer path of neutrons inside the converters. However, it is obvious that for practical reasons the angle cannot be infinitely small and a compromise have to be found between the acceptable value of the detection efficiency on one side and the mechanical constraints and non-uniformity issues due to the transitions between the RPCs on the other.

Another series of simulation was performed for several neutron wavelength (1.8, 2.5, 3.5 and 4.7 Å) at a fixed converter thickness of 3 μm and with the angle ranging from 5º to 15º. As expected from the capture cross section dependence on the neutron energy, the detection efficiency increases with the neutron wavelength. The results presented in figure 10 (right) show that for 4.7 Å neutrons the detection efficiency reaches 68 % at the angle of 5º, demonstrating that $^{10}$B-RPCs at inclined geometry is a promising approach for high detection efficiency PSNDs.

## 4. Conclusions

We have presented a novel thermal neutron PSD concept based on thin-gap RPCs lined with a neutron converter and investigated its feasibility for high detection efficiency and sub-millimeter spatial resolution PSNDs using MC simulations.

While a single $^{10}$B-RPC, irradiated with thermal neutrons at normal incidence has low detection efficiency, this limitation can be overcome by designing the detector with $^{10}$B-RPCs in multilayer or inclined architectures. It is shown that detection efficiencies above 50% can be achieved for both of them by optimizing the neutron converter thickness.

Simulations show that RPCs with thin gas-gaps (≈ 0.35 mm) are better suited for detectors aiming at sub-millimeter spatial resolution than RPCs with wide gas-gaps (≈ 1 - 2 mm). Also, for thin gap RPCs the distribution of the energy deposited by the fission fragments in the gas is significantly narrower, which leads to a lower dynamic range of the induced signals, and, in turn, results in less strict requirements for the front-end electronics.

The obtained results demonstrate that this novel type of PSNDs is particularly well suited for those applications where a high spatial resolution has to be combined with a high accuracy in time, such as, for example, time-resolved or energy-resolved neutron imaging at neutron spallation sources. High detection efficiency, sub-millimeter spatial resolution and sub-nanosecond time resolution, together with the practical advantages such as the simplicity and modularity of the design, robustness and low price per area make RPC-based PSNDs a very promising detection technology for applications in NSS and homeland security.

**ACKNOWLEDGMENTS**

This work was supported in part by the European Union's Horizon 2020 research and innovation programme under grant agreement No 654000.